\documentclass{article}%
\usepackage{amsmath}
\usepackage{graphicx}%
\usepackage{amsfonts}%
\usepackage{amssymb}

\begin{document}

\title{The 8 k - 3 Instanton}
\author{Khaled Abdel--Khalek\\Feza Gursey Institute, P.K. 6, 81220 Cengelkoy, Istanbul, Turkey}
\date{May 2001}
\maketitle

\begin{abstract}
We give an explicit parameterization of the general 8k -3 instanton.

\end{abstract}

\newpage

It seems that gauge field theories have a good chance for describing nature.
At macroscopic and microscopic levels, Abelian and non--Abelian classical and
quantum field theories have an amazing degrees of success in physics or
mathematics. The simplicity and the beauty, manifest and hidden symmetry mix
together in one frame. Non--Abelian gauge theory proved to be a very rich
field of research. Quark confinement stands as a great obstacle to achieve the
first and the oldest dream of high energy physics: What bind nucleons
together? A question that is still rigorously unanswered for decades. For
sure, non--perturbative effects must be taken into account\cite{bpst}.
Constructing all ADHM instantonic solutions explicitly is important
task\cite{adhm}\cite{atiyah}. In this letter, we would like to announce an
explicit full parametrization of instantons solutions in classical gauge field
theory over $R^{2}$ and $R^{4}$.

Initially, we solved the ADHM conditions by iteration%
\[%
\begin{array}
[c]{ccccccccc}%
S^{0} & \longrightarrow & S^{1} & \longrightarrow & \fbox{$\mathit{S}^{3}$} &
\longrightarrow & S^{7} & \longrightarrow & \cdots\\
&  & \downarrow &  & \downarrow &  & \downarrow &  & \\
&  & S^{1} &  & S^{2} &  & S^{4} &  & \cdots
\end{array}
\]
We used the real ADHM over $S^{1}$ (the conformal compactification of $R^{1}$)
to find explicitly the complex ADHM then extending it to quaternions was not
difficult. Here, we will not go into details of the method but we derive the
corresponding $8k-3$ gauge field by means of a simple systematic way that can
be verified easily and compared to the $5k$ multi instanton solutions and at
the same time avoiding many technicalities.

We work over $R^{4}$ which can be endowed with a natural quaternionic
structure%
\begin{equation}
x=x_{\mu}e_{\mu}=x_{0}e_{0}+x_{i}e_{i}\;, \label{xx}%
\end{equation}
for $e_{0}=1$ and $e_{i}e_{j}=-\delta_{ij}+\epsilon_{ijk}e_{k}$, where i,j and
k run from 1 to 3 and $\epsilon_{ijk}$ is the three dimensional skew symmetric
Levi--Civita tensor. We can also represent $e_{0}$ by the $4\times4$
$\operatorname{real}$ identity matrix and $e_{i}$ by the canonical left
$E_{i}$ or right $\overleftarrow{E}_{i}$ quaternionic (for differential
geometric minded readers, we mean hyperKahler) structure\cite{yano}(page 159).
We have the following three $4\times4$ $\operatorname{real}$ antisymmetric
matrices
\[%
\begin{array}
[c]{c}%
(E_{i})_{\mu\nu}=(-\delta_{0\mu}\delta_{i\nu}+\delta_{0\nu}\delta_{i\mu
}-\epsilon_{i\mu\nu})\;,\\
(\overleftarrow{E}_{i})_{\mu\nu}=(-\delta_{0\mu}\delta_{i\nu}+\delta_{0\nu
}\delta_{i\mu}+\epsilon_{i\mu\nu})\;,
\end{array}
\]
explicitly
\[
\left(  E_{i}\right)  _{jk}=-\epsilon_{ijk}\;,\qquad\left(  E_{i}\right)
_{0\nu}=-\delta_{i\nu}\;,\qquad\left(  E_{i}\right)  _{00}=0.
\]
For example%
\[
E_{1}~=~\left(
\begin{array}
[c]{cccc}%
0 & -1 & 0 & 0\\
1 & 0 & 0 & 0\\
0 & 0 & 0 & -1\\
0 & 0 & 1 & 0
\end{array}
\right)  \;,
\]
and so on for the other $E_{i}$ and $\overleftarrow{E}_{i}$. These
quaternionic structures are widely known as the $\eta$ symbols \cite{thooft}.
We can check that%
\begin{align*}
E_{i}E_{j}  &  =-\delta_{ij}+\epsilon_{ijk}E_{k}\;,\\
\overleftarrow{E}_{i}\overleftarrow{E}_{j}  &  =-\delta_{ij}-\epsilon
_{ijk}\overleftarrow{E}_{k}\;.
\end{align*}
One may look at \cite{kh2} and references therein for properties of these $E$'s.

We would like to find the classical $su\left(  2\right)  $ gauge field
$A_{\mu}$ that satisfies the antiself duality equation%
\[
F_{\mu\nu}=-\frac{1}{2}\epsilon_{\mu\nu\alpha\beta}F_{\alpha\beta}\;.
\]
First we introduce the antiself dual $so\left(  4\right)  $ basis
\begin{equation}
\vartheta_{\mu\nu}=\left(  \overline{E}_{\mu}E_{\nu}-\overline{E}_{\nu}E_{\mu
}\right)  \;,
\end{equation}
by $\overline{E}$, we mean conjugation i.e. $\left(  \overline{E}%
_{0}=\overline{E}_{0},\overline{E}_{i}=-\overline{E}_{i}\right)  $, we find
\begin{equation}
\vartheta_{\mu\nu}=-\frac{1}{2}\epsilon_{\mu\nu\alpha\beta}\vartheta
_{\alpha\beta}\;.
\end{equation}
Consider the ansatz%
\begin{equation}
A_{\mu}=-\frac{1}{4}\vartheta_{\mu\nu}\partial_{\nu}\ln\left(  \phi\right)
\ . \label{ans}%
\end{equation}
After simple calculation%
\begin{equation}%
\begin{array}
[c]{c}%
A_{0}=\frac{-\left(  E_{1}\partial_{1}\phi+E_{2}\partial_{2}\phi+E_{3}%
\partial_{3}\phi\right)  }{2\phi}\;,\\
A_{1}=\frac{-\left(  E_{2}\partial_{3}\phi-E_{1}\partial_{0}\phi-E_{3}%
\partial_{2}\phi\right)  }{2\phi}\;,\\
A_{2}=\frac{-\left(  E_{3}\partial_{1}\phi-E_{1}\partial_{3}\phi-E_{2}%
\partial_{0}\phi\right)  }{2\phi}\;,\;\\
A_{3}=\frac{-\left(  E_{1}\partial_{2}\phi-E_{2}\partial_{1}\phi-E_{3}%
\partial_{0}\phi\right)  }{2\phi}\;,
\end{array}
\end{equation}
notice that $A_{\mu}$ is purely imaginary quaternions since it lies in the
su(2) algebra. By straightforward substitution in
\begin{equation}
F_{\mu\nu}=\partial_{\mu}A_{\nu}-\partial_{\nu}A_{\mu}+\left[  A_{\mu},A_{\nu
}\right]  \;,
\end{equation}
``miraculous'' cancellations due to the magic duality leads to
\begin{align}
F_{01}+F_{23}  &  =a_{1}E_{1}+\frac{\partial_{1}\partial_{2}\phi-\partial
_{0}\partial_{3}\phi}{\phi}E_{2}+\frac{\partial_{1}\partial_{3}\phi
+\partial_{0}\partial_{2}\phi}{\phi}E_{3}\nonumber\\
&  =\left(
\begin{array}
[c]{cccc}%
0 & -a_{1} & \frac{-\partial_{1}\partial_{2}\phi+\partial_{0}\partial_{3}\phi
}{\phi} & -\frac{\partial_{1}\partial_{3}\phi+\partial_{0}\partial_{2}\phi
}{\phi}\\
a_{1} & 0 & -\frac{\partial_{1}\partial_{3}\phi+\partial_{0}\partial_{2}\phi
}{\phi} & \frac{\partial_{1}\partial_{2}\phi-\partial_{0}\partial_{3}\phi
}{\phi}\\
\frac{\partial_{1}\partial_{2}\phi-\partial_{0}\partial_{3}\phi}{\phi} &
\frac{\partial_{1}\partial_{3}\phi+\partial_{0}\partial_{2}\phi}{\phi} & 0 &
-a_{1}\\
\frac{\partial_{1}\partial_{3}\phi+\partial_{0}\partial_{2}\phi}{\phi} &
\frac{-\partial_{1}\partial_{2}\phi+\partial_{0}\partial_{3}\phi}{\phi} &
a_{1} & 0
\end{array}
\right)  \;, \label{ff1}%
\end{align}%
\begin{align}
F_{02}+F_{31}  &  =\frac{\partial_{1}\partial_{2}\phi+\partial_{0}\partial
_{3}\phi}{\phi}E_{1}+a_{2}E_{2}+\frac{\partial_{2}\partial_{3}\phi
-\partial_{0}\partial_{1}\phi}{\phi}E_{3}\nonumber\\
=  &  \left(
\begin{array}
[c]{cccc}%
0 & -\frac{\partial_{1}\partial_{2}\phi+\partial_{0}\partial_{3}\phi}{\phi} &
-a_{2} & \frac{-\partial_{2}\partial_{3}\phi+\partial_{0}\partial_{1}\phi
}{\phi}\\
\frac{\partial_{1}\partial_{2}\phi+\partial_{0}\partial_{3}\phi}{\phi} & 0 &
\frac{-\partial_{2}\partial_{3}\phi+\partial_{0}\partial_{1}\phi}{\phi} &
a_{2}\\
a_{2} & \frac{\partial_{2}\partial_{3}\phi-\partial_{0}\partial_{1}\phi}{\phi}
& 0 & -\frac{\partial_{1}\partial_{2}\phi+\partial_{0}\partial_{3}\phi}{\phi
}\\
\frac{\partial_{2}\partial_{3}\phi-\partial_{0}\partial_{1}\phi}{\phi} &
-a_{2} & \frac{\partial_{1}\partial_{2}\phi+\partial_{0}\partial_{3}\phi}%
{\phi} & 0
\end{array}
\right)  \;, \label{ff2}%
\end{align}%
\begin{align}
F_{03}+F_{12}  &  =\frac{\partial_{1}\partial_{3}\phi-\partial_{0}\partial
_{2}\phi}{\phi}E_{1}+\frac{\partial_{2}\partial_{3}\phi+\partial_{0}%
\partial_{1}\phi}{\phi}E_{2}+a_{3}E_{3}\nonumber\\
&  =\left(
\begin{array}
[c]{cccc}%
0 & \frac{-\partial_{1}\partial_{3}\phi+\partial_{0}\partial_{2}\phi}{\phi} &
-\frac{\partial_{2}\partial_{3}\phi+\partial_{0}\partial_{1}\phi}{\phi} &
-a_{3}\\
\frac{\partial_{1}\partial_{3}\phi-\partial_{0}\partial_{2}\phi}{\phi} & 0 &
-a_{3} & \frac{\partial_{2}\partial_{3}\phi+\partial_{0}\partial_{1}\phi}%
{\phi}\\
\frac{\partial_{2}\partial_{3}\phi+\partial_{0}\partial_{1}\phi}{\phi} & a_{3}
& 0 & \frac{-\partial_{1}\partial_{3}\phi+\partial_{0}\partial_{2}\phi}{\phi
}\\
a_{3} & -\frac{\partial_{2}\partial_{3}\phi+\partial_{0}\partial_{1}\phi}%
{\phi} & \frac{\partial_{1}\partial_{3}\phi-\partial_{0}\partial_{2}\phi}%
{\phi} & 0
\end{array}
\right)  \;, \label{ff3}%
\end{align}
where%
\begin{align}
a_{1}  &  =\frac{\partial_{0}\partial_{0}\phi+\partial_{1}\partial_{1}%
\phi-\partial_{2}\partial_{2}\phi-\partial_{3}\partial_{3}\phi}{2\phi}\;,\\
a_{2}  &  =\frac{\partial_{0}\partial_{0}\phi-\partial_{1}\partial_{1}%
\phi+\partial_{2}\partial_{2}\phi-\partial_{3}\partial_{3}\phi}{2\phi}\;,\\
a_{3}  &  =\frac{\partial_{0}\partial_{0}\phi-\partial_{1}\partial_{1}%
\phi-\partial_{2}\partial_{2}\phi+\partial_{3}\partial_{3}\phi}{2\phi}\;.
\label{fgfg}%
\end{align}
Putting together all the conditions arisen from (\ref{ff1}--\ref{fgfg}), we
conclude that \emph{the antiself duality condition holds }\underline{\emph{if
and only if}}%
\begin{equation}
\fbox{$%
\begin{array}
[c]{c}%
\partial_{\mu}\partial_{\nu}\phi=0\;,\\
\partial_{0}\partial_{0}\phi=\partial_{1}\partial_{1}\phi=\partial_{2}%
\partial_{2}\phi=\partial_{3}\partial_{3}\phi\;.
\end{array}
$} \label{as}%
\end{equation}
$F_{\mu\nu}$ simplifies tremendously%
\begin{equation}
F_{\mu\nu}=\frac{-\vartheta_{\mu\nu}\left(  -2\left(  \partial_{0}\partial
_{0}\phi\right)  \phi+\left(  \partial_{0}\phi\right)  ^{2}+\left(
\partial_{1}\phi\right)  ^{2}+\left(  \partial_{2}\phi\right)  ^{2}+\left(
\partial_{3}\phi\right)  ^{2}\right)  }{4\phi^{2}}\;. \label{fmn}%
\end{equation}
Now, we want to solve (\ref{as}), Taking into account that $F_{\mu\nu}$ decays
at infinity with rate $\frac{1}{x^{4}}$. Actually, there are two different
kinds of solutions that we can find: Singular and non--singular. We start by
the \underline{\emph{non--singular}} instanton, the easy solution is what we
call the free part (all the $a$'s are quaternions and $x$ as given in
(\ref{xx}))%
\begin{equation}
\phi_{free}=\lambda_{1}\overline{\lambda}_{1}+\left(  x-a_{1}\right)
\overline{\left(  x-a_{1}\right)  }\;, \label{freeee}%
\end{equation}
after substitution in (\ref{ans}) and (\ref{fmn}), we find directly the $k=1$
BPST instanton\cite{bpst}. Since (\ref{ans}) and (\ref{as}) are additive, we
can construct solutions with higher topological index $k$%
\begin{equation}
\phi_{free}^{k}=\lambda_{1}\overline{\lambda}_{1}+\left(  x-a_{1}\right)
\overline{\left(  x-a_{1}\right)  }+\cdots+\lambda_{k}\overline{\lambda}%
_{k}+\left(  x-a_{k}\right)  \overline{\left(  x-a_{k}\right)  }\;,
\end{equation}
we recover the $5k$ multi--instanton solution. It is not hard to find the
interaction term. From some simple quaternionic reasoning,%
\begin{equation}
\phi_{int}^{12}=\lambda_{1}\overline{\lambda}_{2}+\lambda_{2}\overline
{\lambda}_{1}+\left(  x-a_{1}\right)  \overline{\left(  x-a_{2}\right)
}+\left(  x-a_{2}\right)  \overline{\left(  x-a_{1}\right)  }\ , \label{inact}%
\end{equation}
the important thing to notice is: $\phi_{int}$ is real exactly like
$\phi_{free}$. Unambiguously, we can assemble any $k$ instanton by allowing
all the interacting terms. Let us count the number of parameters: Any
quaternionic $\lambda$ and $a$ contribute $4+4=8$ real parameters, so for
generic $k$ solution, we have $\lambda_{1}\cdots\lambda_{k}$ and $a_{1}\cdots
a_{k}$ quaternions or equivalently $8k$ real parameters. From our familiarity
with the ADHM solution \cite{atiyah}, we know that not all of $\lambda
_{1}\cdots\lambda_{k}$ are independent. The general solution composed of the
free part and the interaction term is invariant under $\lambda\rightarrow
\lambda T$, where $T$ is any generic quaternion of unit norm. Then the true
number of parameters, after substracting the dimension of the automorphism
group, will be $8k-3$ as it should be. Without loss of generality, we can fix
any of the $\lambda$ to be real and the others remain quaternions. for
example, the $k=5$ instanton is%
\begin{align*}
\phi_{full}^{5}  &  =\phi_{free}^{1}+\phi_{free}^{2}+\phi_{free}^{3}%
+\phi_{free}^{4}+\phi_{free}^{5}\\
&  +\phi_{int}^{12}+\phi_{int}^{13}+\phi_{int}^{14}+\phi_{int}^{15}+\phi
_{int}^{23}+\phi_{int}^{24}+\phi_{int}^{25}+\phi_{int}^{34}+\phi_{int}%
^{35}+\phi_{int}^{45}\;,
\end{align*}
with $8\times5-3=37$ real parameters.

We know that the Abelian Seiberg--Witten monopole equations over $R^{4}$
admits only singular solutions. So, we also look for \underline
{\emph{singular}} solutions of the self--dual Yang--Mills instanton. We know
from the index theorem that the most general instanton, whether singular or
non--singular has $8k-3$ real parameters. We can try to plug some minus signs
into (\ref{freeee}) and (\ref{inact}) but we take a different route. We start
by%
\begin{equation}
\phi_{free}=\lambda_{1}\left(  x-a_{1}\right)  \overline{\left(
x-a_{1}\right)  }\bar{\lambda}_{1}\;,
\end{equation}
it satisfies (\ref{as}) and any generic free singular solution can be written
as%
\begin{equation}
\phi_{free}^{k}=\lambda_{1}\left(  x-a_{1}\right)  \overline{\left(
x-a_{1}\right)  }\bar{\lambda}_{1}+\cdots+\lambda_{k}\left(  x-a_{k}\right)
\overline{\left(  x-a_{k}\right)  }\bar{\lambda}_{k}\;,
\end{equation}
finding the interacting term is only possible by allowing mixing between
different indices and at the same time maintaining the reality of $\phi$, then%
\begin{equation}
\phi_{int}^{12}=\lambda_{1}\left(  x-a_{1}\right)  \overline{\left(
x-a_{2}\right)  }\bar{\lambda}_{2}+\lambda_{2}\left(  x-a_{2}\right)
\overline{\left(  x-a_{1}\right)  }\bar{\lambda}_{1}\;.
\end{equation}
In fact singular solutions may be related to the Abelian monopole equation
over $R^{4}$ but the details may take us too far. It is crucial to observe
that the numerator of $F_{\mu\nu}$ (in (\ref{fmn})) is a constant i.e. it has
no $x$ dependence for both of the singular and non--singular solutions.

Returning to the self-duality equation $F_{\mu\nu}=\frac{1}{2}\epsilon_{\mu
\nu\alpha\beta}F^{\alpha\beta}$, we have to use the right instead of the left
quaternionic structure%
\[
E\longrightarrow\overleftarrow{E}\;.
\]
The corresponding $so\left(  4\right)  $ self--dual basis are
\[
\bar{\vartheta}_{\mu\nu}=\left(  \overline{\overleftarrow{E}_{\mu}%
}\overleftarrow{E}_{\nu}-\overline{\overleftarrow{E}_{\nu}}\overleftarrow
{E}_{\mu}\right)  \Longrightarrow\bar{\vartheta}_{\mu\nu}=\frac{1}{2}%
\epsilon_{\mu\nu\alpha\beta}\bar{\vartheta}^{\alpha\beta}.
\]
and proceeding as above, one rederives (\ref{as}).

For the sake of illustration, we conclude with a similar 2 dimensional
problem. We work with u(1) gauge fields (Quantum Electrodynamics QED) over
$R^{2}$ and look for solutions that lead to an integral first Chern class%
\[
\frac{i}{2\pi}\int F_{\mu\nu}dx_{\mu}dx_{\nu}\in\mathbb{Z}\;.
\]
This model in its own is very interesting, it shares many properties of the 4
dimensional problem\cite{polya}. Reducing from 4 dimensions to two dimensions,
we must set to zero all the $e_{2}$ and $e_{3}$ components which amounts to
change quaternions by complex. For example, $x=x_{0}+e_{1}x_{1}$ i.e. only
$E_{1}$ survives
\[
E_{1}=\left(
\begin{array}
[c]{cc}%
0 & -1\\
1 & 0
\end{array}
\right)
\]
and so on. To make clear the discrepancy between the $5k$ and the $8k-3$
solution, we consider explicitly the $k=2$ case, For
\begin{align*}
a_{1}  &  =1+2i\;,\\
a_{2}  &  =5+6i\;,\\
\lambda_{1}  &  =4+0i\;,\\
\lambda_{2}  &  =7+8i\;.
\end{align*}
For the \underline{non--singular} case, we have%

\begin{equation}
\phi_{full}^{2}=285-24x_{0}+4x_{0}^{2}-32x_{1}+4x_{1}^{2}%
\end{equation}
which leads to%
\begin{equation}
F_{01}=\frac{740\;e_{1}}{\left(  285-24x_{0}+4x_{0}^{2}-32x_{1}+4x_{1}%
^{2}\right)  ^{2}}\;.
\end{equation}
Whereas the \underline{singular} case, we find%

\begin{equation}
\phi_{full}^{2}=16\;(151-10x_{0}+7x_{0}^{2}-88x_{1}+7x_{1}^{2})
\end{equation}
and
\begin{equation}
F_{01}=\frac{-904\;e_{1}}{\left(  151-10x_{0}+7x_{0}^{2}-88x_{1}+7x_{1}%
^{2}\right)  ^{2}}\;.
\end{equation}
To make the difference clear, we plot them in Fig. 1 and 2 respectively.

The aim of this article is to provide the general form of instantons in gauge
field theories. But this topological construction is governed by certain
conditions, not every gauge group $G$ over any manifold will lead to
instantons. Over $R^{2}$, it is $\pi_{1}\left(  G=u\left(  1\right)  \sim
S^{1}\right)  =\mathbb{Z}$, and in $R^{4}$, we have $\pi_{3}\left(  su\left(
2\right)  \sim S^{3}\right)  =\mathbb{Z}$ that made possible the construction
of instantons. Hence $u\left(  1\right)  $ in 4 dimensions does not admit
instantonic solutions and the same for $su\left(  2\right)  $ in 6 dimensions
$\left(  \pi_{5}\left(  su\left(  2\right)  =0\right)  \right)  $. Our
$su\left(  3\right)  $ at ten dimensions is quite different. The homotopical
condition is%
\[
Over\;R^{n},\;\;\;\pi_{n-1}\left(  G\right)  =non-trivial.
\]
If instantons have to do anything with confinement then one expects that the
Abelian projection idea should be supplied by some topological considerations.
It will be also interesting to investigate higher dimensional cases.

\bigskip

I gratefully acknowledge the hospitality of the Feza Gursey Institute
especially C. Deliduman, I. H. Duru, O. F. Dayi and C. Saclioglu.

\newpage

\begin{center}%
\begin{figure}
[ptb]
\begin{center}
\includegraphics[
height=3.6434in,
width=4.4979in
]%
{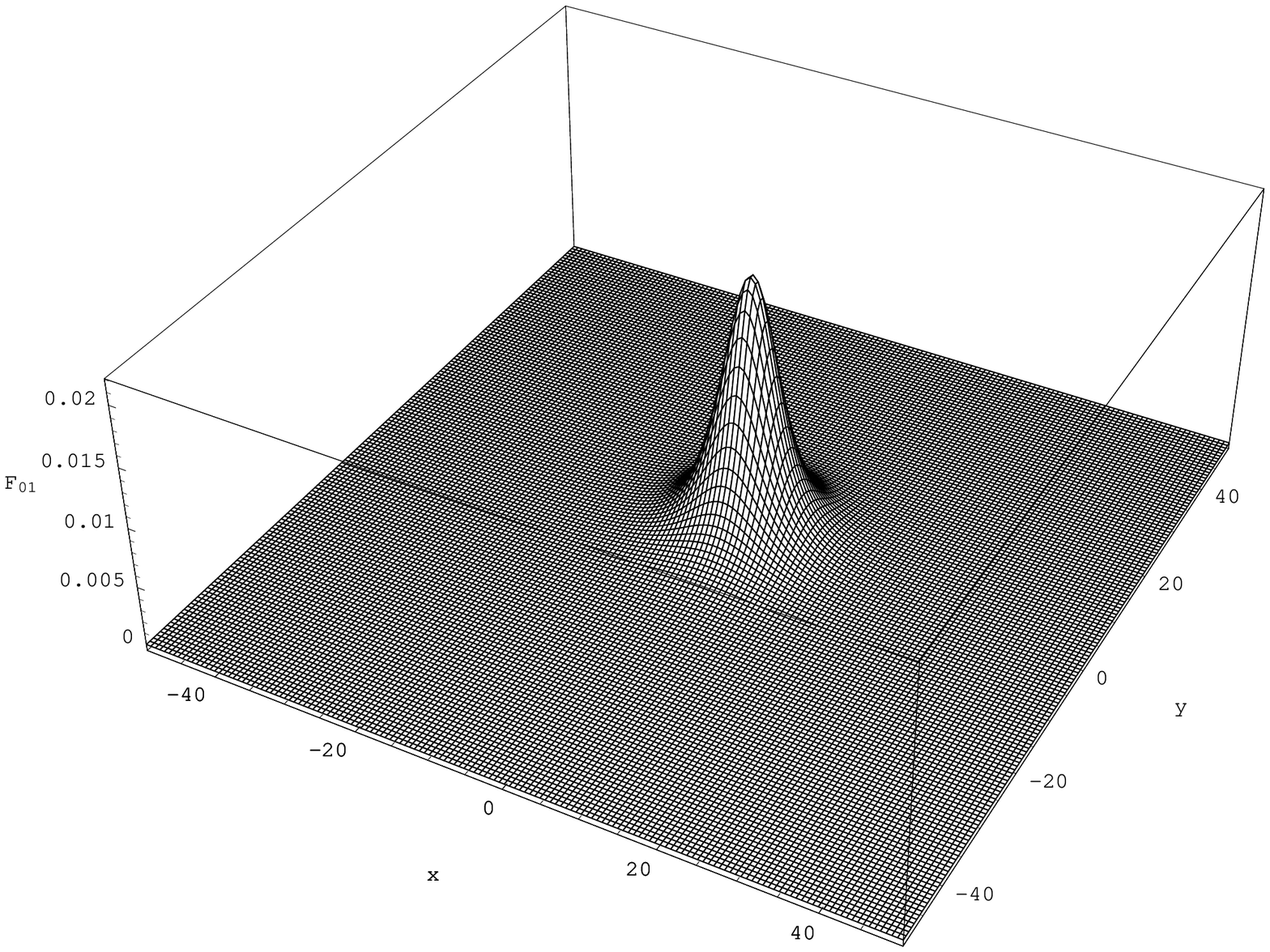}%
\caption{The full non--singular $k=2$ instanton.}%
\end{center}
\end{figure}

\bigskip

\bigskip%
\begin{figure}
[ptb]
\begin{center}
\includegraphics[
height=3.6417in,
width=4.4936in
]%
{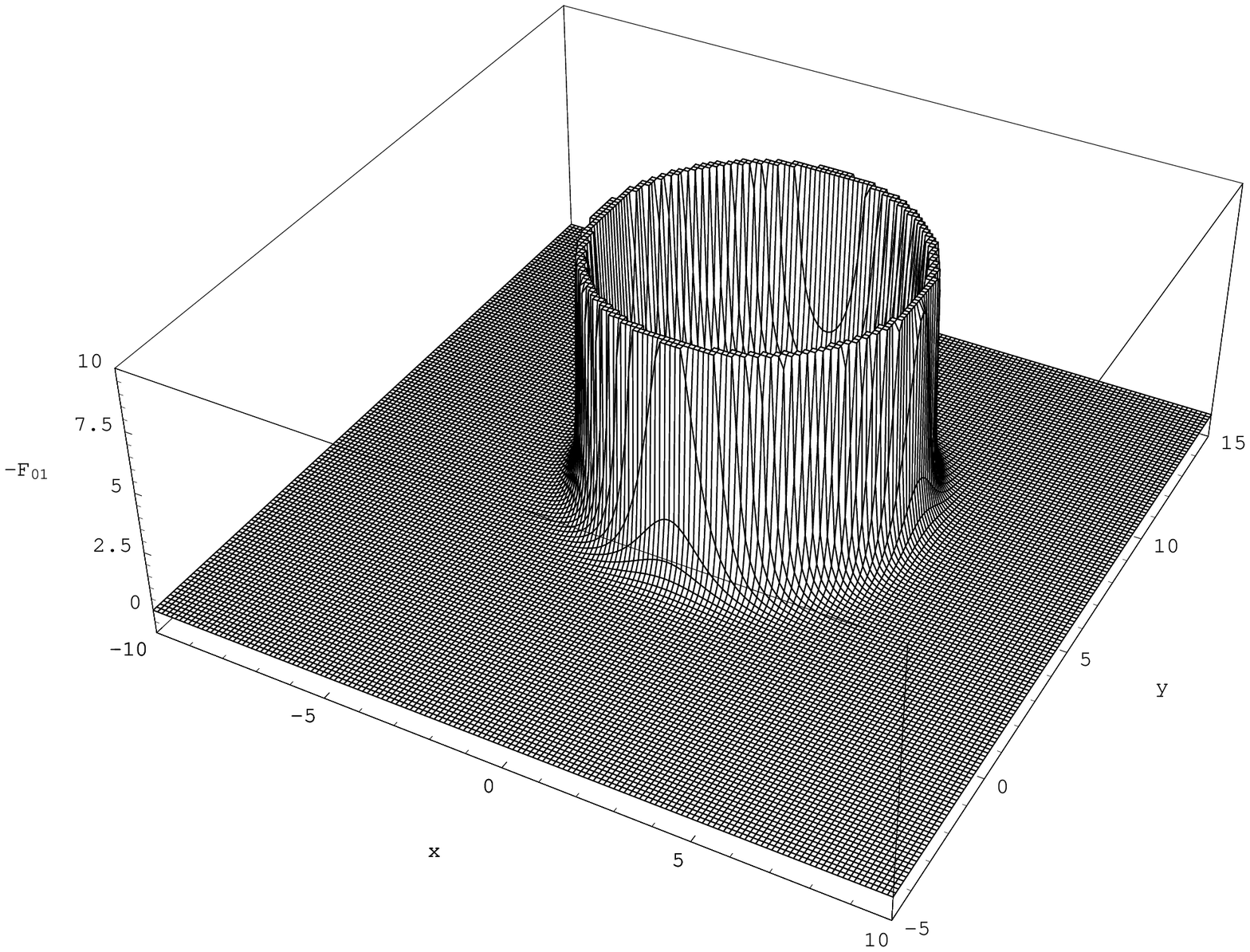}%
\caption{The full singular $k=2$ instanton.}%
\end{center}
\end{figure}

\bigskip

\bigskip
\end{center}
\end{document}